\documentclass[dvips]{article}
\usepackage{icrctc07}

\begin{document}
\title{Interpretation of the atmospheric muon charge
ratio in MINOS}
\shorttitle{charge ratio interpretation}
\authors{Philip Schreiner$^{1,2}$ and Maury Goodman$^1$ for the MINOS collaboration}
\shortauthors{P. Schreiner et al.}
\afiliations{$^1$ Argonne National Lab, $^2$ Benedictine University}
\email{maury.goodman@anl.gov}
\newcommand{\ec}{E_\mu^{surface} \cos \theta}
\newcommand{\esm}{E_\mu^{surface}}
\abstract{
 MINOS is the first large magnetic detector deep underground 
and is the first to measure the muon charge ratio 
with high statistics in the region near 1 TeV.\cite{bib:adamson} An 
approximate formula for the muon charge ratio can be expressed 
in terms of $\epsilon_\pi$ = 115 GeV, $\epsilon_K$ = 850 GeV 
and $\ec$.
The implications for K 
production in the atmosphere will be discussed.
}
\maketitle

\section{Introduction}
Defining A 
\begin{equation}
A \equiv \frac{0.14 E_\mu^{-2.7}}{\rm cm^2~s~sr~GeV}
\end{equation}
the muon energy spectrum underground can be expressed as
\cite{bib:gaisser}:
\begin{equation}
\frac{[dN]}{[dE_\mu]} = A
\left \{ 
\frac{1}
{1+\frac{1.1 E_{\mu}\cos \theta}{\epsilon_\pi}}
  +\frac{0.054}    {1+\frac{1.1 E_{\mu}\cos \theta}{\epsilon_K}}
\right \} 
\end{equation}
where the two terms give the contributions of charged pions
and kaons; the values of $\epsilon = mch_0/\tau$ are close to
the energies where the probability of meson interaction and decay are
equal:  $\epsilon_\pi$ = 115 
GeV and $\epsilon_K$ = 850~GeV.  The zenith angle is $\theta$ and in
experiments with a flat overburden, the largest muon intensity comes at
$\cos \theta = 1$.  
It is possible to generalize this equation to separately
consider $\mu^+$ and $\mu^-$ by introducing constants
$f_\pi$ and $f_K$:

\begin{displaymath}
\frac{[dN^+]}{[dE_\mu]} = A
\left \{ 
\frac{f_\pi}
{1+\frac{1.1 E_{\mu^+}\cos \theta}{\rm 115~GeV}}
  +\frac{0.054 \times f_K}{1+\frac{1.1 E_{\mu^+}\cos \theta}{\rm 850~GeV}}
\right \}
\end{displaymath}

\begin{displaymath}
\frac{[dN^-]}{[dE_\mu]} = A
\left \{ 
\frac{1-f_\pi}{1+\frac{1.1 E_{\mu^-}\cos \theta}{\rm 115~GeV}}
		+  \frac{0.054 \times(1-f_K)}{1+\frac{1.1 E_{\mu^-} \cos \theta}{\rm 850~GeV}}
\right \}
\end{displaymath}
The measured charge ratio in a bin of surface muon energy and zenith angle
can then be expressed as
\begin{eqnarray}
{\normalsize  \frac{N^{\mu^+}}{N^{\mu^-}}  =} 
{\left \{ 
\frac{f_\pi}
{1+\frac{1.1 E_{\mu^+}\cos \theta}{\rm 115~GeV}}
  +\frac{0.054 \times f_K}    {1+\frac{1.1 E_{\mu^+}\cos \theta}{\rm 850~GeV}}
\right \}
}
/\nonumber \\
{
\left \{ 
\frac{1-f_\pi}{1+\frac{1.1 E_{\mu^-}\cos \theta}{\rm 115~GeV}}
		+  \frac{0.054 \times(1-f_K)}{1+\frac{1.1 E_{\mu^-} \cos \theta}{\rm 850~GeV}}
\right \}
}
\label{eq:phil}
\end{eqnarray}
\par An interesting feature of Equation \ref{eq:phil}
is that it depends only on the product $\ec$ and not otherwise
on E or $\theta$.  This combination of terms controls
the relative portions of interaction and decay for both $\pi$'s and K's.
Thus at a fixed value of $\ec$ the ratio of $\mu$'s from $\pi$'s and K's
is constant.  
Equation 3 postulates an
energy independent $\pi^+/\pi^-$ ratio related
to $f_\pi$, an energy independent $K^+/K^-$ ratio related to $f_K$
and an energy independent $\pi/K$ ratio embodied in the 0.054.  These
are all reasonable as a consequence of Feynman scaling.   
(At energies above 10~TeV, charm and the changing
cosmic ray chemical composition
could also affect the muon charge ratio.)
In this simple but powerful parameterization, the charge ratio depends only
on the relative number of $\pi$'s and K's that contribute to muons
measured underground.  
The relative contribution to the muon intensity from
$\pi$'s and K's 
is affected by the relative probability of
hadron interaction and decay,
which vary with both energy and angle.
It varies as
a function of energy because of the hadron lifetimes, and it varies as
a function of angle because of the interaction probability
in the atmosphere.  But as
can be seen from inspection of Equation~\ref{eq:phil}, at any {\it fixed}
value of $\ec$, the ratio of muons from $\pi$ and K decay is also
fixed, even if E or $\theta$
is varied.  

\section{Measurement of $\ec$ underground}
The energy of a muon underground can be related to the energy at the
surface by $E_{surface} = E_{underground} + E_{loss}$ 
The energy loss for muons in MINOS has been calculated
using the energy loss equation, $dE/dx =
a(E) + b(E) \times E$.  Many high
energy muons which
do reach the detector do not bend enough in MINOS' magnetic field to 
measure the curvature and hence charge.  This can be quantified
in terms of the maximum detectable momentum, which depends on the geometry of
the track through the detector \cite{bib:mdm}.  A magnetic detector
underground can measure the charge of muons
that have the lowest energies $E_{underground}$ when they reach the detector.

\par MINOS 
has made measurements of the charge ratio in the far detector
\cite{bib:adamson} which is 710 meters underground or 2070 m.w.e., and
also has yielded a preliminary result using
 its near detector \cite{bib:dejong} which is
100 m underground or 225 m.w.e.  (Note that the far detector is under
a small ridge.)
\par As a crude approximation, we can consider
the case when: 1) the parameter a is constant, 2) the parameter
b is zero. 3) the rock density is constant in all directions, 4)
the surface is flat and 5) the maximum detectable momentum 
is negligible compared to the energy loss from the surface.  
The first four assumptions are equivalent to saying that the
energy loss is proportional to the overburden which only depends
on zenith angle.  In that case $E_{loss} = E_{min}/\cos \theta$
where $E_{min}$ is the minimum energy loss for a vertical cosmic
ray muon.  The fifth assumption implies that
the underground energy of muons used for the charge ratio 
measurement in MINOS was much
smaller than their surface momentum.  The MINOS maximum
detectable momentum, defined at one sigma,
is discussed in these proceedings \cite{bib:mdm}
and a cut was chosen at 2.2 sigma.

In the limit corresponding to these assumptions, the frequency
distribution in 
$\ec$ that we would measure is a delta function at $E_{min}$.
At large zenith angles, the surface energy increases due to
the $\cos \theta$ dependence of the slant depth, but $\ec$ is
constant.
\par
\begin{figure}[h]
\begin{center}
\includegraphics*[width=6cm,angle=0,clip]{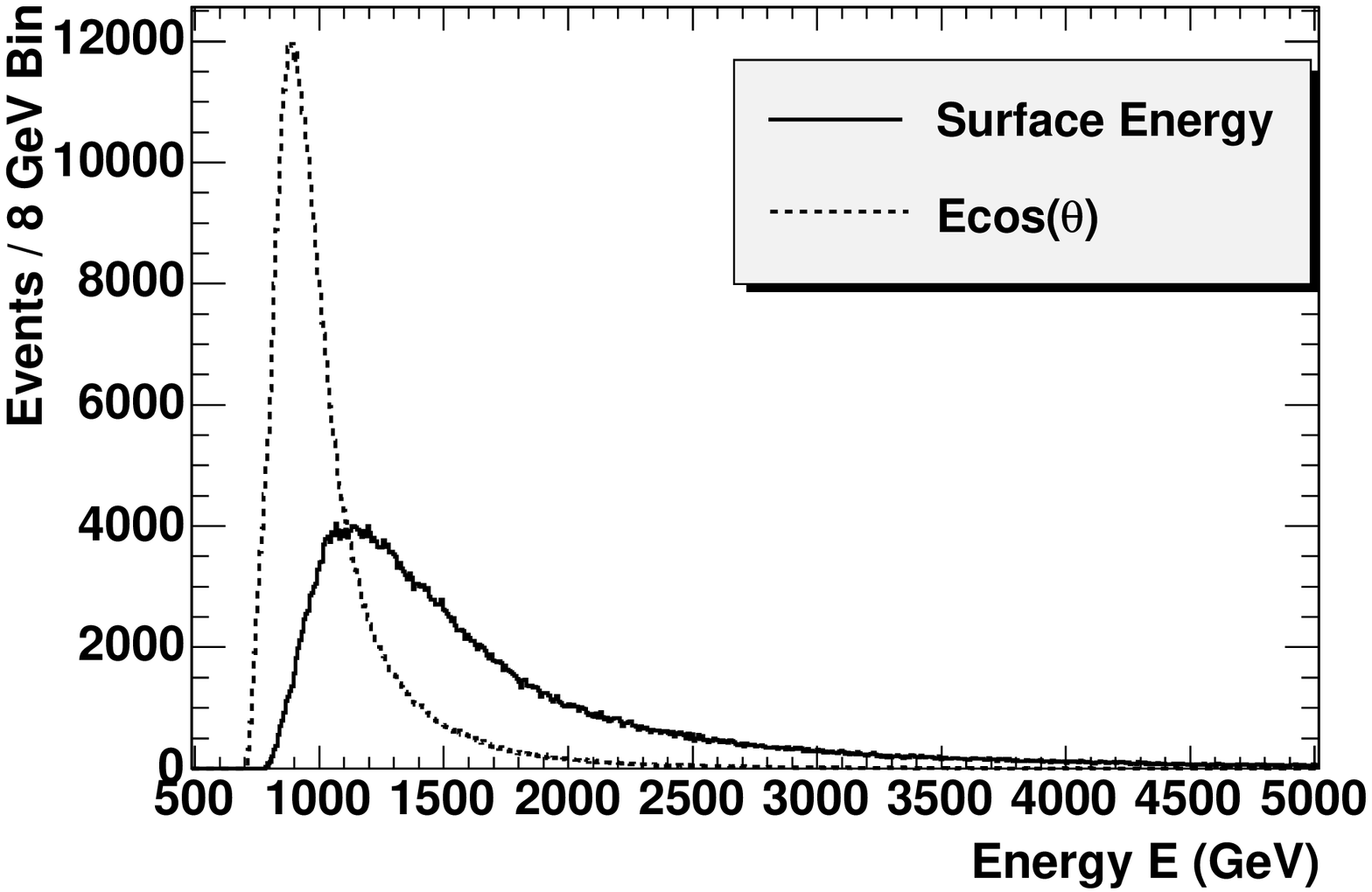}
\caption{\label {fig:farecos} 
Distribution of E and $\ec$ for MINOS data 
muons in the far detector, after cuts.}
\end{center}

\begin{center}
\includegraphics*[width=6cm,angle=0,clip]{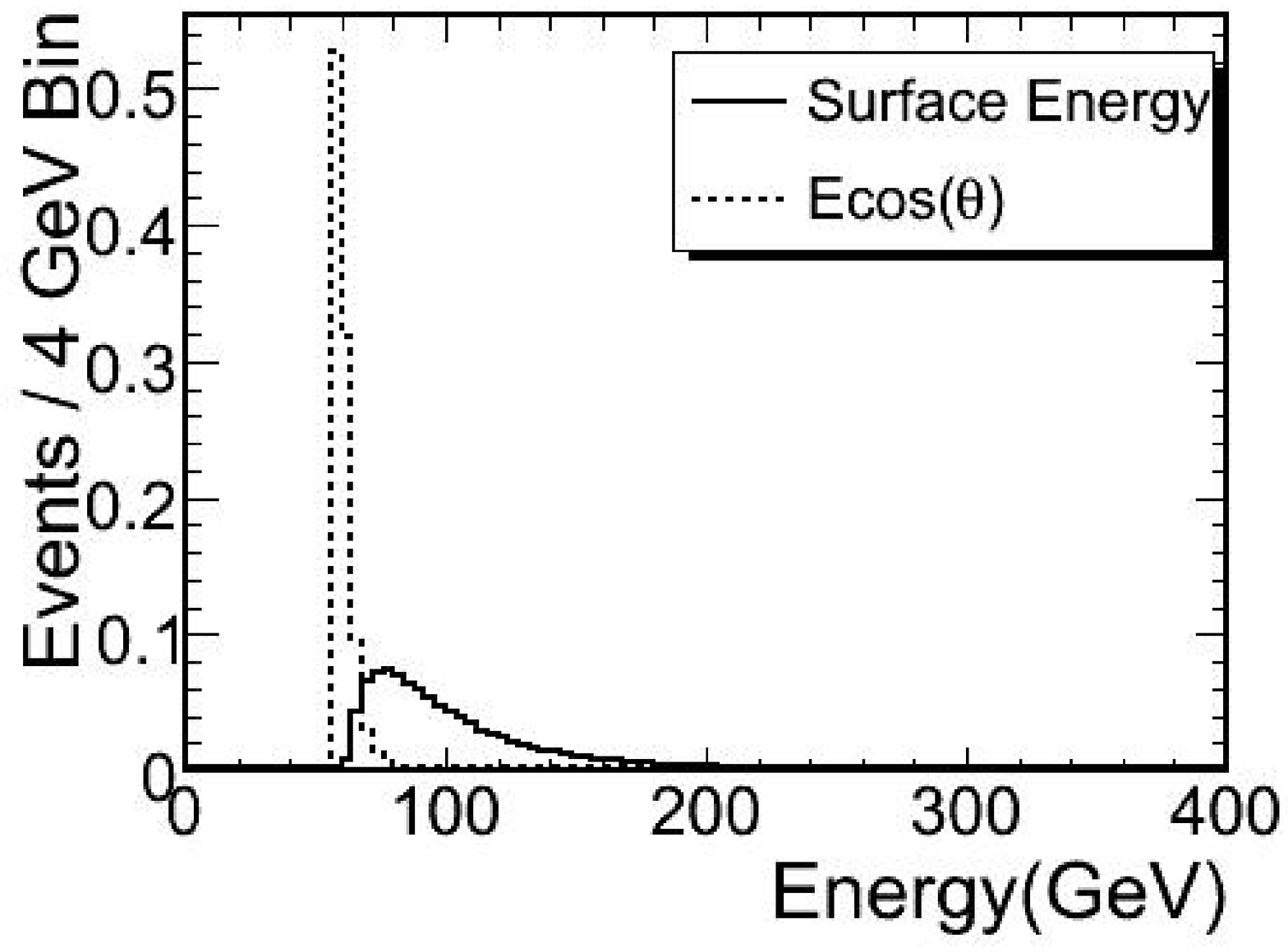}
\caption{\label {fig:nearecos} 
Preliminary distribution of E and $\ec$ for MINOS data muons in the 
near detector, after cuts.}
\end{center}
\end{figure}

This can be compared with the measured E and $\ec$
distributions from MINOS
for the far
detector 
in Figure \ref{fig:farecos} and for a preliminary result
from the near detector in Figure~\ref{fig:nearecos}.
Both $\ec$ distributions are much narrower than the 
corresponding $E_{surface}$ distributions.
The largest contribution to the width of the measured $\ec$ distribution
for the far detector is the b(E) term in the energy loss, while the
largest contribution to the much narrower near detector distribution
is the larger ratio of maximum detectable momentum to energy loss
in the overburden.

\section{Implication for Kaons}
The charge ratio of muons that we measure from $\pi$ decay
is  $r_\pi$ =${\frac{f_\pi}{1-f_{\pi}}}$ 
and from kaon decay $f_K$ =${\frac{f_K}{1-f_K}}$. 
The value 0.054 is related to the relative kaon/pion intensity in 
Gaisser's model.\cite{bib:gaisser}
\par
\begin{figure*}[htbp]
\begin{center}
\includegraphics*[width=11cm,angle=0,clip]{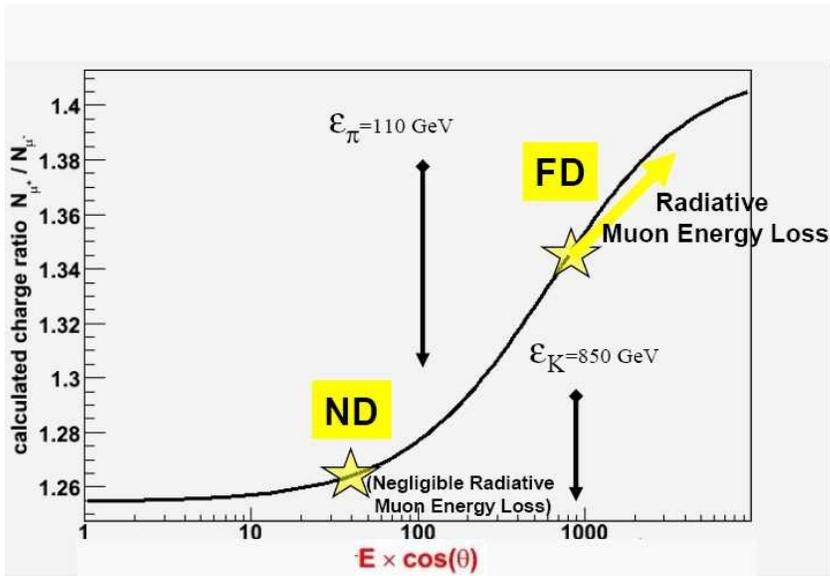}
\caption{\label {fig:peq3} 
Behavior of Equation \ref{eq:phil} for $\ec$ from 1 GeV to 10 TeV.
The critical energies of $\pi$'s 
and $K$'s are shown.  The regions where the MINOS near and far detectors are 
sensitive are also indicated.  Values of $f_\pi$ and $f_K$ from the
MINOS/L3+C fit were used.}
\end{center}
\end{figure*}

A fit was made to MINOS data and the published 
L3+C ratio data \cite{bib:l3} to this parameterization for r.
A chi-square fit was done in 178 bins of $\esm$ vs. $\cos\theta$. The fit 
gave $f_{\pi}= 0.555 \pm 0.002$ and $f_K = 0.667 \pm 0.007$.   The
errors in $f_{\pi}$ and $f_K$ are correlated.
\cite{bib:adamson}
Using these numbers,
the charge ratio versus $\ec$ is plotted in Figure~\ref{fig:peq3}.
 These values for $f_\pi$ and $f_K$ imply that the muon charge ratio from
pion decay is $r=1.25$ and that the muon charge 
ratio from kaon decay is about $r=2.0$.

\begin{figure*}
\begin{center}
\includegraphics*[width=9cm,angle=0,clip]{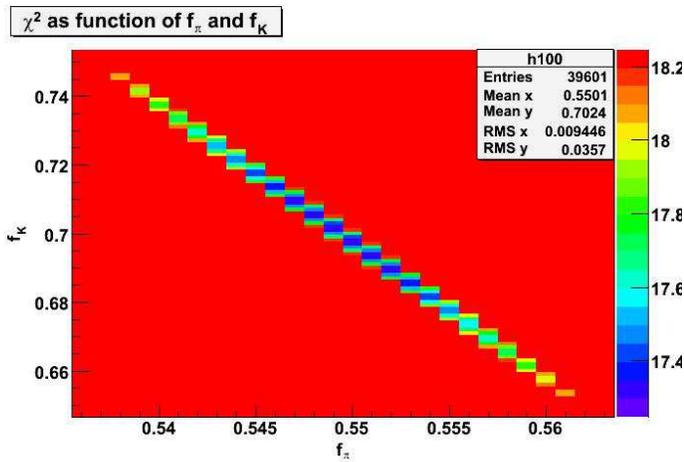}
\caption{\label {fig:fpik} 
Distribution of $\chi^2$ surface in the $f_\pi$ - $f_K$ plane,
showing the error of the MINOS only fit and the correlation
between $f_\pi$ and $f_K$.
}
\end{center}
\end{figure*}

As noted above, there is some $\ec$ variation in the MINOS far
detector data at a value of $\ec$ where the curve in
Figure \ref{fig:peq3} is varying.  
We have performed a fit solely to the MINOS far detector
data, and obtain
$f_\pi$ = $0.549 \pm 0.012$ and $f_K$ = $0.702 \pm 0.049$,
 which imply $r_\pi = 1.22$ and $r_K = 2.36$.
The errors on $f_\pi$ and $f_K$ in this fit are almost completely
correlated and are shown in Figure \ref{fig:fpik}.
A fit which also incorporates the preliminary MINOS near detector charge
ratio
will be presented at the 30th ICRC.

\par Equation \ref{eq:phil} has asymptotic values at both low energy and
high energy.  Note that both values have contributions from both
$\pi$ and K decays.  For the above MINOS-only
fit, the low $\ec$ charge ratio is $r = 1.27$ and the high $\ec$ charge
ratio is $r = 1.45$.
The parameterization
does contain enough physics to (so far) satisfactorily represent
the world's underground muon charge ratio data.  Furthermore, it
is a reasonable way to parameterize the energy dependency
of the charge ratio, and seems preferable to a linear fit in E or
$\log E$.  We suggest that the new MINOS data, equation \ref{eq:phil},
and the fitted values for the model parameters near 1 TeV will 
be useful in the next round of atmospheric neutrino modeling in
this energy range.  It is also now apparent that future precise muon
charge ratio data in the 0.2-0.6 TeV energy range will be useful for
verifying the $\ec$ dependency of the charge ratio.

Our fits to $r_\pi$ give values near 
expectation \cite{bib:naumov}.
Our fit to the $r_K = K^+/K^-$ charge ratio in
atmospheric showers yield values just above 2.  
It is clearly difficult to directly measure this ratio.
Eq. \ref{eq:phil} provides a direction for future study of
this subject.

\section{Acknowledgments}
This work is supported by US Department of Energy 
and National Science Foundation, the UK Particle Physics and 
Astronomy Research Council, and the University of Minnesota.  Operations
of the Fermilab beam are possible only through the large amount of work
of many collaborators and Fermilab employees.
We 
wish to thank the Minnesota Department of Natural Resources for use 
of the facilities of the Soudan Underground State Park, and 
also the large crews of workers who helped construct the detector 
and its components, and the mine crew for help in operating the detector.

\end{document}